\documentclass[12pt]{article}

\usepackage{graphicx}
\usepackage{amscd}

\usepackage{verbatim}

\usepackage{amssymb}

\usepackage{amsmath}

\newcommand{\be}{\begin{equation}}
\newcommand{\ee}{\end{equation}}
\newcommand{\ben}{\begin{eqnarray}}
\newcommand{\een}{\end{eqnarray}}
\newcommand{\ba}{\begin{eqnarray}}
\newcommand{\ea}{\end{eqnarray}}

\newcommand{\bi}{\begin{itemize}}
\newcommand{\ei}{\end{itemize}}

\begin{document}

\begin{center}

\vspace{24pt} { \large \bf The sausage sigma model revisited} \\

\vspace{30pt}

\vspace{30pt}

\vspace{30pt}

{\bf Vardarajan
Suneeta}\footnote{suneeta@iiserpune.ac.in}

\vspace{24pt} 
{\em  The Indian Institute of Science Education and Research (IISER),\\
Pune, India - 411008.}

\end{center}
\date{\today}
\bigskip

\begin{center}
{\bf Abstract}
\end{center}
Fateev's sausage sigma models in two and three dimensions are known to be integrable. We study their stability under RG flow in the target space by using results from the mathematics of Ricci flow. We show that the three dimensional sausage is unstable, whereas the two dimensional sausage appears to be stable at least at leading order as it approaches the sphere. We speculate that the stability results obtained are linked to the classification of ancient solutions to Ricci flow (i.e., sigma models which are nonperturbative in the IR regime) in two and three dimensions. We also describe a class of perturbations of the three dimensional sausage (with the same continuous symmetries) which remarkably decouple. This indicates that there could be a new solution to RG flow which is described at least perturbatively as a deformation of the sausage.

\newpage
\section{Introduction}
Two dimensional nonlinear sigma models arise in string theory as the quantum field theory (QFT) on the world-sheet traced by a closed string in a target-space. The metric of the target space is the coupling constant in the sigma model, and flows when we do a renormalization group (RG) transformation of the sigma model. The fixed points of the RG flow are exact string backgrounds and the world-sheet sigma models in these cases are conformal field theories (CFTs) --- a study of these CFTs is naturally of interest in string theory. From the QFT point of view, there are several subjects of investigation in sigma model RG flow - for example, the (in)stability of fixed points on the space of solutions to the RG flow, the end-point of possible instabilities, and whether there are any attractors on the space of solutions, to name a few.
Sigma model RG flow is a geometric flow on the target-space manifold, and these questions can be posed geometrically as well. Issues such as stability of fixed points can be discussed using two complementary approaches, either in terms of deformations of the world-sheet CFT or geometrically, in the target space. Recently, it has been conjectured that solutions to RG flow could approximately describe dynamical processes in string theory (see \cite{GMT} for a review). Explicit solutions to RG flow have been found that are conjectured to describe the process of closed string tachyon condensation and it has been shown for these examples that spacetime energy decreases under RG flow \cite{APS}, \cite{GHMS}. Algorithms using the simplest RG flow, the Ricci flow (a subject of active study in mathematics), have been developed, to find non-trivial solutions to the vacuum Einstein equation \cite{headrick}. The stability of Ricci-flat fixed points under RG flow have been shown to be linked to their stability in quantum gravity \cite{suneeta}.

The simplest RG flows are target space metrics of constant curvature which either expand or contract uniformly under the RG flow truncated to first order. A natural next step given an RG flow is to understand the ultraviolet (UV) and infra-red (IR) limits of the world-sheet QFT. The hyperbolic target space expands uniformly under the flow and the IR limit of the flow is a free field theory. On the other hand, the sphere contracts uniformly under the first order flow and when its curvature becomes sufficiently large, the first order flow is no longer valid --- nonperturbative effects become significant.
It is not possible to study this regime in the target space picture through a geometric flow truncated to some order.

For the target space a two-sphere (i.e, $O(3)$ sigma model), Fateev, Onofri and Zamolodchikov (henceforth, FOZ) \cite{foz} investigated the IR limit of the sigma model. The IR limit of the RG flow of $O(3)$ sigma models depends crucially on the topological terms in the model. For certain choices, the world-sheet QFT is integrable, and even in such cases, the IR limit can either be a field theory with a mass gap or a CFT depending on the nature of the topological terms. The ultraviolet (UV) limit, on the other hand, is trivial --- the curvature of the sphere decreases in the UV limit, and the sigma model is asymptotically a free field theory. In addition to having nontrivial UV and/or IR limits, in general, RG flow geometries can become singular. However, for compact two dimensional target spaces, there are powerful results on the formation of singularities under first-order RG flow (the  Ricci flow). These results imply that if the target-space is of genus zero, then any initial metric flows to the sphere --- the sigma model approaches the IR limit of the $O(3)$ sigma model. The sphere/$O(3)$ sigma model is thus an attractor on the space of solutions to RG flow in this case. Similarly, a metric on a genus one compact surface flows to the flat torus metric, and a metric on a higher genus surface flows to the hyperbolic metric. In particular, no type of singular behaviour is possible for compact surfaces, other than approaching the shrinking round sphere.

The nonperturbative IR behaviour of the sphere/$O(3)$ sigma model is reminiscent of Yang-Mills theory. The fact that the nonperturbative behaviour is completely understood from the work of FOZ means that such field theories could serve as toy models for understanding Yang-Mills theory better. The attractor behaviour of this sigma model implies that the IR behaviour of a whole class of sigma models which flow to the sphere in the IR can be obtained just from the nonperturbative IR limit of the $O(3)$ model. These observations lead to some natural questions: (i) Are there other sigma models with nonperturbative IR behaviour? (ii) Is the IR limit of these field theories well-understood? (iii) Are some of these attractors?

There is another sigma model for which the answer to the first two questions is in the affirmative.
 This is an RG flow on a two dimensional genus zero compact target space --- the sausage sigma model --- first obtained by FOZ \cite{foz} and is an integrable field theory. The target-space is a sausage which gets rounder and shrinks along the flow, approaching the sphere (the sigma model is nonperturbative in the IR and approaches the IR limit of the $SO(3)$ model). In this regime, the sausage sigma model can be viewed as an integrable deformation of the $O(3)$ sigma model. It is therefore \emph{completely understood} in all regimes.

Mathematicians have also been interested in solutions to Ricci flow which are non-singular in the UV but singular in the IR --- these are called \emph{ancient} solutions. The interest in ancient solutions from the point of view of physics is that, as we described, they are toy models to understand Yang-Mills theory. A recent classification result in mathematics implies that there are only two such ancient solutions in two dimensions, the two-sphere and the sausage \cite{sesum}. Any other sigma model on genus zero compact target spaces is also nonperturbative in the UV.

There is a generalization of some of these results to three dimensional target spaces. In particular, Fateev found a three dimensional `sausage' solution \cite{fateevs3}, \cite{fateevs3paper2} which is non-singular in the UV, and approaches the IR limit of the $O(4)$ sigma model (shrinking three-sphere). It is an ancient solution of the flow, and the corresponding three dimensional sausage sigma model is integrable \cite{lukyanov}. Hence the sigma model is well-understood in all regimes. It is also a special case \cite{tseytlin} of a wider class of integrable models discussed in \cite{klimcik1}, \cite{klimcik2}.  However, unlike two dimensional target spaces, the first-order RG (Ricci) flow in three dimensions is far more complicated. For example, starting with an initial metric even on a simply connected compact target space, the flow can lead to neck-like singularities. The sausage is simply connected and compact, but has no neck-like singularities --- as it is an ancient solution, its only singular behaviour (under Ricci flow) is approaching the shrinking sphere. A natural question is if there are other such ancient solutions approaching the $O(4)$ sigma model in the IR. It is known that there is at least one more rotationally symmetric ancient solution other than the sausage and the three sphere on compact, simply connected three dimensional target spaces \cite{perelman} although the precise metric is not available (a construction of the solution by a limiting procedure is given in \cite{chow}). A complete classification of ancient solutions in three dimensions is still lacking. Ancient solutions have been constructed on higher dimensional target spaces in analogy with Fateev's sausage in \cite{bakas}. Ancient solutions to RG flow with a $B$ field have been obtained in \cite{lukyanov}. Several interesting common features of integrable sigma models, and of the sausage sigma models in particular, have been explored recently from a field theory viewpoint \cite{lukyanov1}, \cite{lukyanov2}.

What we propose in this paper is a geometric study of the sausage models in the target space. Specifically, the sausage solutions have many features in common with the spheres, such as nonperturbative IR behaviour and integrability.
Two and three dimensional spheres are also attractors for nearby solutions --- i.e., they are `geometrically stable' under the RG flow. Is this true for the sausages in two and three dimensions? If they were attractors as well, they would be the model QFTs for IR behaviour of a whole class of sigma models which approached them in the IR. This would be desirable as the sausage sigma models are well-understood in all regimes. Another issue we would like to explore is whether there is a \emph{distinct geometric feature} (such as geometric stability) associated with all ancient solutions \footnote{On compact target spaces which are simply connected, we do not know of any (Ricci flow) solutions which are not ancient but integrable sigma models.}. In particular, we would like to investigate geometric stability of the sausage solutions (in a linearized approximation) in two and three dimensions to determine if they are attractors.

The plan of the paper is as follows: in section 2, we introduce the two dimensional sausage and some of its geometric features. We also study its perturbations and explain our notion of linear stability for the sausage. We repeat this analysis for the three dimensional sausage in section 3. Surprisingly, we find differences in the stability properties of the sausage in two and three dimensions, implying that geometric stability is not a feature of every integrable sigma model. In section 4, we describe a class of trace-free perturbations of the three dimensional sausage which remarkably decouple, indicating that there could be new ancient solutions (to Ricci flow) generalizing those considered by Fateev (but with the same continuous symmetries). Finally, in the Discussion section, we summarize our results and speculate on the reasons for the difference in the stability properties of the sausage in two and three dimensions.

\section{Perturbing the two dimensional sausage}
It is well-known that the RG flow of the target-space metric of the world-sheet sigma model (to first order in the squared string length $\alpha'$) is the Ricci flow, a well-studied equation in mathematics,
\begin{eqnarray}
\frac{\partial g_{ij} }{\partial t} = - 2 R_{ij},
\label{3.2}
\end{eqnarray}
where the RG flow parameter $\tilde t = \frac{2 t}{\alpha'}$.
If, along the flow, curvature becomes large (of order $\alpha'$) then this truncation to first order in $\alpha'$ is invalid and in general, it is not possible to study the RG flow perturbatively.

Consider this flow on a two dimensional surface, whose metric is of the form
\begin{eqnarray}
ds^2 = u(x, \theta, t) [dx^2 + 4 d\theta^2].
\label{3.3}
\end{eqnarray}

Thus, the metric is conformal to the metric on a cylinder of radius $2$, $ds^{2}_{cyl} = [dx^2 + 4 d\theta^2]$.

In this case, Ricci flow (\ref{3.2}) reduces to the flow of the conformal factor $u(x,\theta, t)$, given by
\begin{eqnarray}
\frac{\partial u }{\partial t} = \Delta_{cyl}  [ \ln u ] ;
\label{3.4}
\end{eqnarray}
where $\Delta_{cyl}$ is the Laplacian operator on the cylinder of radius $2$, $\Delta_{cyl} = \frac{\partial^2}{\partial x^2} + \frac{\partial^2}{4 \partial \theta^2 } .$

A solution to the flow (\ref{3.4}) with axial symmetry is the conformal factor of the sausage sigma model discussed by Fateev, Onofri and Zamolodchikov (FOZ henceforth) \cite{foz};
\begin{eqnarray}
u_{0}(x,t) =\left ( \frac{1}{\lambda} \right ) \frac{\sinh(-\lambda t) }{ \cosh x + \cosh (\lambda t)}.
\label{3.5}
\end{eqnarray}

$\lambda >0$ is a positive real parameter, and this metric extends smoothly to the compactified surface obtained by adding two points to the two cylinder ends --- hence the name `sausage' for this geometry. In the UV limit (as $t \to - \infty$), the sausage becomes longer and can be visualized as two cigar shapes (two Euclidean Witten black holes \cite{witten}) joined together at their ends. This statement is very precise, as it can be shown that the neighbourhood of either of the sausage tips in the backward limit $t \to - \infty$ reduces to the Euclidean Witten black hole metric. This is done by a coordinate scaling of the sausage metric before taking the backward limit, to `zoom in' on the tip (the sausage can be viewed as an asymptotics-changing perturbation of the Witten black hole \cite{cigarsuneeta}).

The metric is defined for $-\infty <t < 0$ and the curvature becomes singular in the limit as $t \to 0$ --- indicating that the IR limit of the sausage sigma model is nontrivial and nonperturbative effects become important.
For any non-zero finite $t$, the surface has the characteristic sausage shape that becomes rounder and rounder as $t \to 0$. The evolution of the sausage under the flow is shown schematically in Figure 1.
\begin{figure*}
\centerline{\mbox{\includegraphics[width=2.20in]{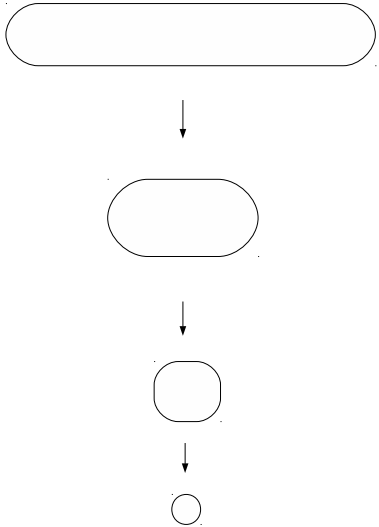}}}
\caption{Flow of the sausage}
\label{fig1}
\end{figure*}
Mathematically, as $t \to 0$, the sausage metric is given by the round sphere metric of radius $\sqrt{2t}$ at leading order followed by corrections proportional to higher powers of $t$. There is a (stronger) field-theoretic version of this observation --- as shown by FOZ, the sausage sigma model which is integrable can in fact be viewed (as $t \to 0$) as an integrable deformation of a class of $O(3)$ sigma models. The IR limit of these $O(3)$ sigma models is discussed by FOZ and can either be a field theory with a mass gap or a CFT depending on the topological terms in the sigma model. The $O(3)$ sigma model is an attractor on the space of solutions to RG flow, and the sphere is geometrically stable under Ricci flow.

Following the discussion in the Introduction, we would like to investigate if the sausage solution itself is geometrically stable under Ricci flow at least in a linearized analysis. We perturb the sausage geometry and study the evolution of the perturbation under the RG flow (\ref{3.2}). Working in conformal gauge, our metric is still of the form (\ref{3.3}) where now $u(x, \theta, t) = u_{0}(x,t) + h(x, \theta, t)$. Since $h$ is a perturbation, we can consider the flow of $u(x, \theta, t)$ given by (\ref{3.4}) in a linearized approximation for $h$. This yields the linearized flow of $h$;
\begin{eqnarray}
\frac{\partial h }{\partial t} = \Delta_{cyl}  \left ( \frac{h}{u_{0}} \right ).
\label{3.6}
\end{eqnarray}

Clearly, this is not an easy equation to solve since $u_{0}$ is not separable (as a function of $x$ and $t$). There is another way to view this non-separable solution. The flow of $u(x,t)$ is a particular case of the class of fast diffusion equations, and arises in plasma physics. In this context, the sausage solution $u_{0}(x,t)$ was discussed by Rosenau \cite{rosenau} and is known in Ricci flow literature as the Rosenau solution. As $t \to- \infty$, we can visualize the conformal factor $u_{0}$ as two widely separated kinks which come closer as $t$ increases and coalesce into one pulse, which shrinks and vanishes as $t \to 0$. Any perturbation of the two-kink system is hard to analyze when they are separated. However, $u_{0}(x,t)$ can be approximated by a separable function of $x$ and $t$ in the phase when the kinks have coalesced into one pulse. (\ref{3.6}) should be tractable in this regime. This corresponds to the limit $t \to 0$ when the sausage geometry approaches the sphere.
Further, results from mathematics \cite{hamilton} imply that the metric on any compact genus zero surface must approach the sphere under Ricci flow. So, in fact, any perturbation of the sausage must also approach the sphere.

Therefore, we perturb the sausage as it approaches the round sphere (i.e., as $t \to 0$) and compare the rates at which both the perturbed geometry and the sausage approach the round sphere metric. We will make this more precise soon.  We have
\begin{eqnarray}
u_{0}(x,t) = - \frac{t}{2}~ \text{sech}^2 (\frac{x}{2}) -  \lambda^2 t^3 ~ \text{sech}^2 (\frac{x}{2}) \left ( \frac{1}{6} - \frac{\text{sech}^2 (\frac{x}{2})}{4} \right ) + ......,
\label{3.7}
\end{eqnarray}
where the ellipses refer to terms with higher powers of $t$. Thus, for small $t$, retaining only the first term, we have
$u_0 \sim - \frac{t}{2}~ \text{sech}^2 (\frac{x}{2}) $ and the next order correction is proportional to $t^3$.
Retaining only the leading term in $u_{0}$ in (\ref{3.7}) amounts to the approximation that the background metric is that of the round sphere. This can be explicitly seen by a coordinate transformation $\tanh \frac{x}{2} = \cos \beta$, under which we recover a round sphere of radius $\sqrt{2(-t)}$  labelled by angles $\beta$ and $\theta$. Thus $|g^{sausage}_{ij} - g^{sphere}_{ij}| \sim O(t^3)$.

If we plug only the leading term in (\ref{3.7}) back into (\ref{3.6}), and look for a separable solution $h(x, \theta, t)$,
we obtain
\begin{equation}
h(x, \theta, t) = \text{sech}^2 \left( \frac{x}{2}\right )~ P_{l}^{m} (\tanh \frac{x}{2})~ e^{im\theta}~(-t)^{2C} ;
\label{separablesoln}
\end{equation}
where $m$ is a real constant; and $P_{l}^{m}$ are the associated Legendre functions. Also, $4C = l(l+1)$. For a nonsingular solution, we take $l$ and $m$ to be nonnegative integers with $0 \leq m \leq l$. The appearance of associated Legendre functions is not surprising since at leading order in $t$, perturbing the sausage is like perturbing the round sphere.

 We will now explain how we can investigate the stability of the sausage this way. Powerful results from the mathematics of Ricci flow imply that any $n$-sphere is geometrically stable under the flow --- i.e., up to gauge, it is an attractor for nearby solutions \cite{hamilton}, \cite{hamilton1}, \cite{huisken}. A separate result on linear stability of the sphere has also been proven in \cite{ilmanen},\cite{caozhu} using a monotonic functional under this flow \cite{perelman1}. The upshot of all this is that any perturbation of the sausage must approach the sphere as well, and at leading order, the sausage metric is the sphere metric. The rate of approach of the perturbed sausage to the sphere is important. Denoting the perturbed metric components by $g^{perturbed}_{ij}$, we are interested in finding $p$ such that $|g^{perturbed}_{ij} - g^{sphere}_{ij}| \sim O(t^p)$ as $t \to 0$. If $p<3$, the perturbed metric would approach the sphere \emph{slower} than the sausage does, and we would conclude the sausage is not an attractor for nearby solutions.

Let us show that $|g^{perturbed}_{ij} - g^{sphere}_{ij}| \sim O(t^3)$, i.e., $p=3$. For the separable solution (\ref{separablesoln}), the $t$ dependence is $ \sim (-t)^{2C}$ , with $4C = l(l+1)$. When $l=0$, we have a constant in $t$ solution which will dominate our background metric. When $l=1$, we have a solution which has a linear dependence on $t$ and would be of the same order as our background metric, therefore not a perturbation. Both these $l$ values are ruled out as we already know that the sphere is geometrically stable under the flow. Therefore these modes correspond to spurious solutions which will not extend to solutions of Ricci flow.  When $l=2$, we see that $C = 3/2$ and this perturbation is of order $t^3$. This is therefore the leading order behaviour of the perturbation, and there is no order $t^2$ term. Of course, to obtain this, we have retained only the leading order term in $u_0$ in (\ref{3.7}). We expect that the order $t^3$ perturbation should in general receive contributions from the order $t^3$ term in the expansion of $u_0$ in (\ref{3.7}) as well. Let us assume that as $t \to 0$, the perturbation (\ref{separablesoln}) and its corrections in higher powers of $t$ can be expanded as follows:
 $$ h (x, \theta, t) = \text{sech}^2 \left( \frac{x}{2}\right ) \left [ f_1(x, \theta) t^3 + f_2 (x, \theta) t^{3 + \alpha}
 + ..... \right ];$$
where $\alpha > 0$ and ellipses denote higher powers of $t$. This is reasonable, as $u_0$ can be expanded in powers of $t$ as shown in (\ref{3.7}). If we substitute this back in (\ref{3.6}), then we see that there is no contribution to the order $t^3$ term in $h(x, \theta, t)$ from next-to-leading-order terms in $u_0$. They only contribute at order $t^5$.

Due to the absence of the order $t^2$ term in the perturbation of the sausage, we have established that arbitrary (even non-rotationally symmetric) perturbations of the sausage approach the sphere at the same rate as the sausage itself, or faster. There do exist non-zero perturbations which approach the sphere at the same rate as the sausage, and are given by $l=2$ in (\ref{separablesoln}). If they extend to solutions of the full RG flow, our stability result would be akin to marginal stability in dynamical systems. However, these are the only such perturbations --- \emph{all} other perturbation modes with $l > 2$ evolve as $t^{3+\alpha}$ where $\alpha > 0$, and do not affect the rate of approach of the sausage to the sphere. We will see in the next section that these results do not generalize to the case of the sausage in three dimensions.
\section{The three-dimensional sausage geometry and its perturbations}

We now review the three-dimensional generalization of the sausage solution. This was proposed by Fateev \cite{fateevs3}, \cite{fateevs3paper2}. This is a solution to the RG flow (\ref{3.2}). Its geometric properties have been discussed in \cite{bakas}, and we will use the notation given in that paper to rewrite the RG flow as
\begin{eqnarray}
\frac{\partial g_{ij} }{\partial \tau} = \frac{1}{2} R_{ij},
\label{4.1}
\end{eqnarray}
where the RG flow parameter $\tilde t = -\frac{\tau}{2 \alpha'}$. Since increasing $\tilde t$ corresponds to decreasing $\tau$, studying the RG flow to the IR corresponds to integrating (\ref{4.1}) backwards in $\tau$. The UV limit corresponds to $\tau \to \infty $. In terms of $\tau$, the three dimensional sausage geometry (parametrized in terms of $\theta \in [ 0, \frac{\pi}{2} ]$,
$\chi_{1}, \chi_{2} \in [0, 2\pi ]$) can be written as
\begin{eqnarray}
ds_{\nu}^{2} = A(\tau, \theta)d\theta^2 + B(\tau, \theta)d\chi_{1}^{2} + C(\tau, \theta)d\chi_{2}^{2};
\label{4.2}
\end{eqnarray}
where, letting $\xi = \nu \tau$,
\begin{eqnarray}
A(\tau, \theta) &=& \frac{1}{\nu} \frac{\cosh \xi \sinh \xi }{(\cos^2 \theta + \sin^2 \theta \cosh \xi)(\sin^2 \theta + \cos^2 \theta \cosh \xi)}, \nonumber \\
B(\tau, \theta) &=& \frac{1}{\nu}\frac{\cos^2 \theta \sinh \xi }{(\sin^2 \theta + \cos^2 \theta \cosh \xi )}, \nonumber \\
C(\tau, \theta) &=& \frac{1}{\nu}\frac{\sin^2 \theta \sinh \xi }{( \cos^2 \theta + \sin^2 \theta \cosh \xi )}.
\label{4.3}
\end{eqnarray}

This metric has a generalization to include a term of the form $2D(\tau, \theta) d\chi_1 d\chi_2 $ with a specific $D(\tau, \theta)$ as discussed in \cite{fateevs3paper2}, \cite{bakas} --- inserting this metric into the sigma model, it was shown by Fateev that (even with $D \neq 0$) the sigma model is integrable. This is similar to the case of the sausage sigma model in the previous section. There are more similarities between the two solutions. When $\tau \to 0$, this metric approaches the metric of a shrinking three-sphere. As in the sausage model whose metric approaches that of a shrinking two-sphere, the IR limit is therefore nontrivial. In these coordinates, as $\tau \to 0$, we have the limiting metric $$ds^2 = \tau ( d\theta^2 + \cos^2 \theta d\chi_{1}^{2} + \sin^2 \theta d\chi_{2}^{2} ).$$
This corresponds to a three-sphere of radius $\tau$ in Hopf coordinates, the sigma model for which has $O(4)$ symmetry. Thus, for the metric (\ref{4.3}) or its generalization with $D \neq 0$, only translations of $\chi_1 $ and $\chi_2$ correspond to isometries, and the corresponding sigma model only has $U(1) \times U(1)$ symmetry.

Henceforth, for the sake of computational simplicity, we will not consider a nonzero $D$ term, i.e., we will analyze only the metric (\ref{4.3}). This metric has one other feature which is similar to the 2d sausage --- as shown in \cite{bakas}, upon change of coordinates, scaling of a coordinate and then taking the UV limit $\tau \to \infty$, the metric approaches a product of the Euclidean Witten black hole metric with a circle. Thus we call this, by analogy, the `three-dimensional sausage'.

We would now like to consider small perturbations of this geometry and evolve the perturbations under the flow (\ref{4.1}).
For this, we first note that RG flow is not form-invariant under diffeomorphisms that depend on the flow parameter. The RG flow is modified by terms depending on the generator of the diffeomorphisms $V_{i}$. In terms of the parameter $\tau$, the modified flow (in the new coordinates) is
\begin{eqnarray}
\frac{\partial g_{ij}}{\partial \tau} = \frac{1}{2}
(R_{ij} + \nabla_{i} V_j + \nabla_{j} V_i )
\label{4.4}
\end{eqnarray}
The choice of $V_j$ corresponds to `fixing gauge' and we can choose this to simplify the flow equation in specific cases.
Let us consider $g_{ij}^{P} = g_{ij} + h_{ij}$ where $g_{ij}$ now denotes the $3d$ sausage metric, and $h_{ij}$ is a perturbation of this geometry small enough in magnitude for a linearized approximation to be valid. Then, applying this approximation to (\ref{4.4}), the flow of $h_{ij}$ is
\begin{eqnarray}
\frac{\partial h_{ij}}{\partial \tau} = - \frac{1}{4} [ -
(\Delta_{L}  h)_{ij} + \nabla_{i}\nabla_{j} H -  \nabla_{i}(
\nabla^{c}h_{cj}) - \nabla_{j}( \nabla^{c}h_{ci})  + 2
\nabla_{i} V_j + 2  \nabla_{j} V_i  ].~~~~~
\label{4.5}
\end{eqnarray}

Since we work in the linearized approximation, all covariant derivatives are taken with
respect to the background metric $g$. $H = g^{ij} h_{ij}$ is the
trace of the perturbation.
\begin{eqnarray}
(\Delta_{L} h)_{ij} = - \Delta h_{ij} + 2 R_{~ijc}^{d} h_{~d}^{c} +
 R_{i}^{c} h_{jc} +  R_{j}^{c} h_{ic} \label{4.6}
\end{eqnarray}
is the Lichnerowicz laplacian acting on symmetric 2-tensors (all
curvature tensors are those of the three dimensional sausage metric). The
convention we follow for the Lichnerowicz laplacian is that of the
physics literature, and differs from the mathematics one by a
negative sign. $\Delta h_{ij} = g^{kl}\nabla_{k}\nabla_{l}h_{ij}$ is the usual Laplacian acting on 2-tensors.

Since we have the freedom to choose $V_i $ as we like, we will choose it so as to get rid of divergence terms in (\ref{4.5}).
This leads to the simplified form
\begin{eqnarray}
\frac{\partial h_{ij}}{\partial \tau} = - \frac{1}{4} [ -
(\Delta_{L}  h)_{ij}].
\label{4.7}
\end{eqnarray}
We note that we need to integrate the above equation backwards in $\tau$ as the RG flow parameter $\tilde t = -\frac{\tau}{2 \alpha'}$. This is not an easy task, as the covariant derivatives and the curvature tensors in (\ref{4.6}) are all of the three dimensional sausage metric (\ref{4.3}) which itself depends on $\tau$ (thus we cannot separate variables). Recent results from Ricci flow on simply connected three-manifolds (for example, any type of perturbed three dimensional sausage) imply that either the geometry can develop neck-like singularities, or the metric will approach that of the shrinking three-sphere \cite{perelman1}. Thus our perturbed three dimensional sausage could develop curvature singularities. If so, then even though the geometry in the singular portions is known from the mathematics literature, this is a nonperturbative regime in RG flow, and it is not possible for us to make any statements about the associated quantum field theory. Thus, we will only consider the other possibility, namely that the perturbed sausage approaches the three-sphere metric, as does the (unperturbed) sausage. We can confine ourselves in the regime $\tau \to 0$ and compare the rates of approach of the sausage and the perturbed sausage to the round three-sphere. We want to determine $p$ where $|g^{perturbed}_{ij} - g^{sphere}_{ij}| \sim O(\tau^p)$ as $\tau \to 0$. It can be checked from (\ref{4.3}) that  $|g^{sausage}_{ij} - g^{sphere}_{ij}| \sim O(\tau^3 )$.
Thus, if $p < 3$ then the sausage cannot be an attractor for nearby solutions as the perturbed geometry approaches the sphere slower than the sausage. Our goal is therefore to find $p$.

The problem is more complicated than the two dimensional case. While studying the perturbations of the two-dimensional sausage, we only considered the flow of the trace of the perturbation as we could work in conformal gauge in two dimensions. Here, we will need to study the evolution of the trace-free part of the perturbation as well.

First, let us consider the flow of the trace of the perturbation in order to draw parallels with the two dimensional case. Under Ricci flow, trace of a perturbation is not `gauge', hence must be studied. Let the trace of the perturbation be $H = g^{ij}h_{ij}$. Tracing the equation (\ref{4.7}), we get
\begin{eqnarray}
g^{ij} \frac{\partial h_{ij}}{\partial \tau} = - \frac{1}{4} [ \Delta H ].
\label{4.8}
\end{eqnarray}

$\Delta H$ is the usual scalar Laplacian. This can be rewritten as
\begin{eqnarray}
\frac{\partial H}{\partial \tau} = - \frac{1}{4} [ \Delta H ] - \frac{1}{2} h_{ij} R^{ij},
\label{4.9}
\end{eqnarray}
where we have used the fact that given the flow (\ref{4.1}), $\frac{\partial g^{ij} }{\partial \tau} = - \frac{1}{2} R^{ij}.$

Now, as $\tau \to 0$, to leading order, the background metric is that of a 3-sphere of radius $\sqrt{\tau}$. As in the two dimensional case, we will see that this leading order term in the background metric is enough to give us the exponent $p$, such that $|g^{perturbed}_{ij} - g^{sphere}_{ij}| \sim O(\tau^p)$. We will also use the result referred to in the earlier section, that the $n$-sphere is geometrically stable under the Ricci flow. Nearby solutions approach the 3-sphere (up to gauge).

Then, at leading order in $\tau$,  $R^{ij} = \frac{2}{\tau} g^{ij}$, and $\Delta H  = \frac{1}{\tau} \bar \Delta H $ where $\bar \Delta H$ is the Laplacian on the unit 3-sphere. Choosing $H$ to be a spherical harmonic we have ($l = 1,2,.....$)
\begin{eqnarray}
\tau \frac{\partial H}{\partial \tau} &=&  - \frac{1}{4} [\bar \Delta H ] - H; \nonumber \\
&=& [\frac{l(l+2)}{4} - 1]H.
\label{4.10}
\end{eqnarray}

Let us denote the metric components on a unit 3-sphere as $\bar g_{ij} $. Then $H = g^{ij}h_{ij} = \frac{1}{\tau} \bar g^{ij} h_{ij}$. Given the geometric stability of the 3-sphere, we know that $h_{ij}$ has to vanish faster than the leading order background geometry. This requires that $h_{ij} \sim \tau^{1 + \epsilon }$, where $\epsilon > 0$. This in turn implies $H \sim \tau^{\epsilon }$. From (\ref{4.10}) we see that for $l = 1$, $H \sim \tau^{-1/4} $ --- we do not consider this further as the perturbed geometry has to approach the 3-sphere of radius $\sqrt{\tau}$. The $l=1$ mode will therefore not extend to a full solution of Ricci flow. For $l = 2$, $H \sim \tau$, which implies that there are at least some components of the perturbation $h_{ij} \sim \tau^2 $. Thus, for these perturbations, $p=2$. This shows that the three dimensional sausage behaves very differently from the two dimensional version. In particular, the perturbed sausage approaches the 3-sphere slower than the sausage. A simple example of this class of perturbations is $h_{ij} = f(x^k )g_{ij}$.

We can consider more general classes of perturbations by taking advantage of the fact that as $\tau \to 0$, the background metric to leading order is that of a 3-sphere. Recall that on a 3-sphere, a symmetric tensor can be decomposed in a standard way in terms of various harmonics on the sphere:
$$ h_{ij} = h^{TT}_{ij} + \nabla_i W_j + \nabla_j W_i + H g_{ij} + \nabla_i \nabla_j F - \frac{1}{3} g_{ij} \Delta F ;$$

where $h^{TT}_{ij}$ is transverse ($\nabla^{i} h^{TT}_{ij} = 0$) and trace-free ($g^{ij}h^{TT}_{ij} = 0$) and described by a basis of tensor spherical harmonics;  $W_j$ is transverse ($\nabla^{i} W_{i} = 0$) and can be expressed in a basis of vector spherical harmonics; and $H$, the trace of the perturbation, and $F$ can be expressed in terms of scalar spherical harmonics.

 We can use this decomposition to obtain the leading order in $\tau$ behaviour of the perturbation as that can be obtained with the background sausage metric being approximated by the sphere metric. In fact, we can restrict to trace-free metrics (the trace already having been studied) and further restrict to each class of trace-free perturbations --- those which are transverse-traceless, the ones proportional to vector spherical harmonics, and the ones proportional to the scalar spherical harmonics. This decomposition cannot be used to obtain next-to-leading-order corrections in $\tau$ for $h_{ij}$ as these classes would all couple in general at next order. Indeed, note that
 for consistency of the flow (\ref{4.7}), we require that if $h_{ij}$ is transverse-traceless (TT), $(\Delta_{L} h)_{ij}$ must also have this property. This is not true in general --- it is only true of backgrounds whose curvature tensor obeys a specific condition (a derivation of the condition is given, for example, in p.9, \cite{suneeta}). In particular, if the background metric is Einstein, this property holds and one can consistently restrict to TT perturbations (Theorem 4.6, \cite{besse}). Therefore, for the sausage, the best we can do is restrict to perturbations which are trace-free - this property being preserved by the Lichnerowicz laplacian $\Delta_L $ for any background. We can further demand that the perturbations be  `TT to leading order' as $\tau \to 0$. By this we mean that $\nabla^{i} h_{ij}$ is not zero, however, considering the covariant derivative evaluated using the 3-sphere metric , and the leading order piece in $h_{ij}$ as $\tau \to 0$, $\nabla^{i} h_{ij}$ is zero at leading order. Then, at this order, the perturbation can be expanded in a basis of TT tensor spherical harmonics on the 3-sphere. The spectrum of the Lichnerowicz laplacian on these tensors (on the \emph{unit} 3-sphere) is known:
$$(\Delta_L h)_{ij} = [l(l+ 2) - 2] + 6   ,$$
for integers $ l \geq 2$. We have the background metric of a 3-sphere of radius $\sqrt{\tau}$ at leading order in $\tau$, and therefore, the flow (\ref{4.7}), for such perturbations is ($h^{TT}_{ij}$ is a tensor harmonic):
\begin{eqnarray}
\tau \frac{\partial h^{TT}_{ij}}{\partial \tau} = \frac{[l(l+2) + 4] }{4} h^{TT}_{ij};~~~~~~~ l \geq 2.
\label{4.11}
\end{eqnarray}
For $l=2$, we have $h_{ij} \sim \tau^3 $. Thus perturbations in this class do not approach the 3-sphere slower than the sausage.

If $h_{ij} = \nabla_i W_j + \nabla_j W_i $ at leading order ($W_j $ a vector harmonic), then
\begin{eqnarray}
\tau \frac{\partial h_{ij}}{\partial \tau} = \frac{[l(l+2) + 1] }{4} h_{ij};~~~~~~~ l \geq 1.
\label{4.12}
\end{eqnarray}
For $l=1$, $h_{ij} \sim \tau$ and therefore will not extend to a solution to Ricci flow. For $l=2$, we have $h_{ij} \sim \tau^{9/4}$ which definitely approaches the sphere slower than the sausage.

If $h_{ij} = c \nabla_i \nabla_j F + \frac{1}{3} g_{ij} F $ at leading order ($F$ is a scalar harmonic, and choose $c$ so that this perturbation is trace-free),
then
\begin{eqnarray}
\tau \frac{\partial h_{ij}}{\partial \tau} = \frac{[l(l+2)] }{4} h_{ij};~~~~~~~ l \geq 0.
\label{4.13}
\end{eqnarray}
In this case, $l=0,1$ have to be discarded, but for $l=2$, we have $h_{ij} \sim \tau^2 $.

These are some examples of special perturbations for which the spectrum of the Lichnerowicz laplacian is known for the 3-sphere \cite{higuchi}, \cite{ishikodama}. There are thus many classes of perturbations for which the perturbed sausage approaches the 3-sphere slower than the sausage. The sausage in three dimensions cannot be an attractor for nearby solutions. This behaviour is markedly different from two dimensional case. In the final section, we will speculate on the possible reasons.

\section{A special class of tracefree perturbations}

We have seen in the previous section that we can consistently restrict to sub-classes of perturbations of the sausage such as TT perturbations only at leading order as $\tau \to 0$. In this section, we identify a special sub-class of trace-free perturbations which remarkably decouple for all $\tau$.
These are perturbations for which the only nonzero components are $h_{\theta \chi_1}, h_{\theta \chi_2}$. These perturbations are obviously tracefree. Furthermore, we require that $h_{\theta \chi_1}$ and $h_{\theta \chi_2}$ are functions only of $\theta$. Thus the perturbed metric is invariant under translations of $\chi_1$ and $\chi_2$, like the sausage metric.

We show that there are nontrivial solutions to the flow (\ref{4.7}) in this class. In the Appendix, we have derived the components of $(\Delta_L h)_{ab}$ in detail for the three dimensional sausage metric valid for all $\tau$ ($a,b$ refer to any of the three coordinate indices). For the special class of perturbations considered in this section, the only components of $(\Delta_L h)_{ab}$ that are nonzero are $(\Delta_L h)_{\theta \chi_1}$ and $(\Delta_L h)_{\theta \chi_2}$ (see the Appendix for the details). Furthermore, the flow (\ref{4.7}) reduces to two decoupled partial differential equations for $h_{\theta \chi_1}$ and $h_{\theta \chi_2}$. We use (\ref{a10}) in the Appendix to obtain the flow equation for
$h_{\theta \chi_1}(\theta, \tau)$  (primes refer to partial derivatives with respect to $\theta$)  :
\begin{eqnarray}
\frac{\partial h_{\theta \chi_1} }{\partial \tau} &=& \frac{1}{4A} \left [ -\partial_{\theta}^2 h_{\theta \chi_1}
+ \left ( \frac{3 A'}{2A} + \frac{B'}{2B} - \frac{C'}{2 C} \right ) \partial_{\theta} h_{\theta \chi_1}
+ \right. \nonumber \\&& \left.  \left ( \frac{A''}{2A} - \frac{(A')^2 }{A^2 } - \frac{3 B''}{2B} + \frac{3 (B')^2 }{2 B^2 } + \frac{A' B'}{2 A B}
+ \frac{(C')^2 }{2 C^2 } + \frac{A' C' }{2 A C} - \frac{C''}{2C} \right ) h_{\theta \chi_1} \right ].~~~~~~~~~~
\label{5.1}
\end{eqnarray}
This equation is to be integrated in backwards $\tau$. The physical RG flow parameter $\tilde t = -\frac{\tau}{2 \alpha'}$. The flow equation for $h_{\theta \chi_2}$ is similar to (\ref{5.1}) --- it is obtained by interchanging $B$ and $C$ and of course, replacing $h_{\theta \chi_1}$ with $h_{\theta \chi_2}$.
This differential equation is an exact equation for this class of perturbations in the three dimensional sausage background. It is not separable since $A$, $B$, $C$ are nonseparable functions of $\tau$ and $\theta$. This makes it difficult to solve, even as $\tau \to 0$. In principle, of course, numerical methods could be used to integrate the equation, at least as $\tau \to 0$. But it is indeed surprising that we have this set of decoupled equations involving only two perturbation components. The perturbed spacetime has the same continuous symmetries as the sausage. We end this section with two questions: is this indicative of a new ancient solution generalizing Fateev's solutions \cite{fateevs3} and \cite{fateevs3paper2}? Is it integrable?

\section{Discussion} In this paper, we have studied the stability of the two and three dimensional sausage solutions of Fateev under the appropriate first-order RG flow (the Ricci flow). The two and three dimensional sausage sigma models are integrable deformations of the (integrable) $O(3)$ (2-sphere) and $O(4)$ (3-sphere) sigma models respectively. Geometrically, the sausages approach the spheres (under the first-order flow) as the RG flow parameter $t$ approaches a finite value (which we can take to be zero). The rate at which the perturbed sausage approaches the sphere is compared with the rate at which the sausage approaches the sphere. This is used to obtain a measure of stability at leading order in $t$ as $t \to 0$.

The stability problem of the sausage is motivated by the fact that it has many geometric features in common with the sphere. The $n$-sphere is additionally stable under perturbations and an attractor for nearby solutions to RG flow. Attractors are model QFTs for the IR behaviour of all sigma models that approach them, hence this is a very desirable geometric feature. The question therefore is if this is a feature of the sausage as well, and for all solutions for which the sigma model is integrable. If true, this would also establish a link between a geometric property (stability) and a world-sheet property (integrability). We find the result to be in the negative. The perturbed two dimensional sausage approaches the 2-sphere at the same rate as the sausage itself. However, in three dimensions, for a wide class of perturbations (for which we have given examples in section III), the perturbed sausage approaches the sphere \emph{slower }than the sausage. This indicates that the sausage in three dimensions is not an attractor for nearby solutions, whereas in two dimensions, it appears to be stable at least at leading order in $t$. How do we understand the stability result for the $n$-sphere which is already known, and what we have found for the sausages?
In the terminology of Ricci flow, the $n$-sphere is a \emph{Ricci soliton} --- which is any solution to Ricci flow changing at most by an overall scale and/or gauge. The stability properties of various Ricci solitons have been analyzed by \cite{ilmanen} and \cite{caozhu} and not all Ricci solitons are stable. Besides, the two dimensional sausage, which is not a Ricci soliton, appears to be stable at leading order in $t$ as $t \to 0$. We speculate that the stability results obtained are linked to the classification of ancient solutions in two and three dimensions. In two dimensions, on compact, genus zero surfaces, we have only two possibilities - the 2-sphere and the 2-sausage \cite{sesum}. In three dimensions, we do not have a complete classification of ancient solutions but we know there is at least one more in addition to the sausage and the 3-sphere \cite{perelman}. Could there be a perturbation of the 3-sausage that in fact approaches a third ancient solution? Could this be one possible unstable mode of the 3-sausage approaching the 3-sphere slower than the sausage? These are questions we hope to investigate in future.

In section IV, we explicitly demonstrated that there exist a class of nontrivial perturbations of the three-sausage with only two non-zero components under the linearized flow (\ref{4.7}). The two components obey two decoupled flow equations making them amenable to study beyond the leading order. The perturbed geometry has the same continuous symmetries as the three-sausage. The decoupling of these perturbations is surprising and could be perturbative evidence of a new solution to the RG/Ricci flow generalizing Fateev's examples. From the point of Ricci flow, the interesting questions are, is there a new solution which in the perturbative regime is described by these two non-zero components? Is it a new ancient solution? From the point of physics, it would be interesting to study this perturbation in a world-sheet picture as a deformation of Fateev's sausage. The question to investigate there would be if this deformation is also integrable.

\section{Appendix: Useful formulae}

Here, we put together useful formulae related to the three dimensional sausage metric (\ref{4.2}). In what follows, primes refer to derivatives with respect to $\theta$. Recall that $A,B,C$ are functions only of $\tau$ and $\theta$ and are given by (\ref{4.3}). Note: \emph{in some formulae, repeated indices are not summed over. Therefore, the convention for repeated indices will be explicitly mentioned before each set of formulae.} Primes refer to partial derivatives with respect to $\theta$.
Nonzero Christoffel symbols are:
\begin{eqnarray}
\Gamma_{\theta \theta}^{\theta} &=& \frac{A'}{2A};~ \Gamma_{\chi_1 \chi_1}^{\theta} = \frac{-B'}{2A};~ \Gamma_{\chi_2 \chi_2}^{\theta} = \frac{-C'}{2A}; \nonumber \\
\Gamma_{\theta \chi_1}^{\chi_1} &=& \frac{B'}{2B};~\Gamma_{\theta \chi_2}^{\chi_2} = \frac{C'}{2C}.
\label{a1}
\end{eqnarray}

The non-zero Ricci tensor components are:
\noindent
\begin{eqnarray}
R_{\theta \theta} &=& \frac{1}{4} \frac{A'}{A} \left [\frac{B'C + C'B}{BC} \right ] + \frac{1}{4} \frac{(B')^{2}}{B^2} + \frac{1}{4} \frac{(C')^{2}}{C^2} - \frac{1}{2} \frac{B''}{B} - \frac{1}{2} \frac{C''}{C}; \nonumber \\
R_{\chi_1 \chi_1} &=& - \frac{B''}{2A} + \frac{1}{4} \frac{B' A'}{A^2 } - \frac{1}{4} \frac{B' C'}{AC} + \frac{1}{4} \frac{(B')^{2}}{A B} ; \nonumber \\
R_{\chi_2 \chi_2} &=& - \frac{C''}{2A} + \frac{1}{4} \frac{C' A'}{A^2 } - \frac{1}{4} \frac{C' B'}{AB} + \frac{1}{4} \frac{(C')^{2}}{AC}.
\label{a2}
\end{eqnarray}
\vskip 0.5 cm
The Lichnerowicz Laplacian acting on the perturbation, from (\ref{4.6}), and for the three dimensional sausage metric is given by the formulae below:\\
\\
\noindent
(a) $i,j $ indices refer to $\chi_1$ or $\chi_2$ and repeated $i,j$ indices are summed in the formula below over $\chi_1$ and $\chi_2 $:
\begin{eqnarray}
(\Delta_L h)_{\theta \theta} = - (\Delta h)_{\theta \theta} - 2(g_{ij}h^{ij}) R_{\theta \theta} -2 g_{\theta \theta} R_{ij} h^{ij} + R (g_{ij} h^{ij})g_{\theta \theta} + 2 R_{\theta \theta} h^{\theta \theta}.~~~~~
\label{a3}
\end{eqnarray}
\\
\noindent
(b) We next write $(\Delta_L h)_{ii}$ where $i$ is either $\chi_1 $ or $\chi_2 $. Therefore, in the formula below, when the $i$ index is repeated, it is NOT summed. The index $j \neq i \neq \theta$. Therefore there is no summation over any repeated indices.
\begin{eqnarray}
(\Delta_L h)_{ii} &=& - (\Delta h)_{ii} - 2 g_{ii} R_{\theta \theta} h^{\theta \theta} -2 g_{\theta \theta} h^{\theta \theta} R_{ii} +
R g_{ii}g_{\theta \theta} h^{\theta \theta} \nonumber \\ &&- 2 g_{jj} h^{jj} R_{ii} - 2 g_{ii} R_{jj}h^{jj} + R g_{jj}g_{ii}h^{jj}
+ 2 g^{ii} R_{ii} h_{ii}.~~~~
\label{a4}
\end{eqnarray}
\\
\noindent
(c) With the same conditions on indices as in the previous equation (no summation on repeated indices, and $i$ is either $\chi_1 $ or $\chi_2 $, $j \neq i \neq \theta$ ),
\begin{eqnarray}
(\Delta_L h)_{\theta i} = - (\Delta h)_{\theta i} + 2 g^{\theta \theta} R_{\theta \theta} h_{\theta i} + 2 g^{ii} R_{ii} h_{\theta i} - g^{jj} R_{jj} h_{\theta i}.~~~
\label{a5}
\end{eqnarray}
\\
\noindent
(d) Finally,
\begin{eqnarray}
(\Delta_L h)_{\chi_1 \chi_2 } = - (\Delta h)_{\chi_1 \chi_2 } + \left [ 3 g_{\chi_1 \chi_1} R_{\chi_2 \chi_2 } + 3 g_{\chi_2 \chi_2} R_{\chi_1 \chi_1 } - R g_{\chi_1 \chi_1}g_{\chi_2 \chi_2} \right ] h^{\chi_1 \chi_2} .~~~~~~~~~
\label{a6}
\end{eqnarray}
\\

Next, we list lengthy expressions for the various Laplacians appearing in (\ref{a3}--- \ref{a6}).
The partial derivative of any Christoffel symbol with respect to $\chi_1$ or $\chi_2$ is zero for the sausage background as the sausage metric components are functions only of $\theta$ and $\tau$. This can be used to simplify the formulae we have listed even further. Indeed, we will also give the value of each of the Laplacians for a special class of perturbations for which the only nonzero components are $h_{\theta \chi_1} (\theta, \tau)$ and $h_{\theta \chi_2} (\theta, \tau)$.\\

1. In the formula below, $i$ index is for $\chi_1$ or $\chi_2$ and sum over the $i$ index is shown explicitly - in this case, summation is over $\chi_1$ and $\chi_2 $:
\begin{eqnarray}
&&(\Delta h)_{\theta \theta} = g^{\theta \theta} \partial_{\theta}^{2} h_{\theta \theta} - g^{\theta \theta} \partial_{\theta} \left [ \frac{A'}{A} h_{\theta \theta} \right ] -
\frac{3A'}{2A} \partial_{\theta}h_{\theta \theta} + \frac{3}{2} \left (\frac{A'}{A} \right )^2 h_{\theta \theta} +  \nonumber \\ && \sum\limits_{i=\chi_1,\chi_2} \left [ g^{ii}\partial_{i}^{2} h_{\theta \theta} - 2 g^{ii} \partial_i (\Gamma_{i\theta}^{i}) h_{i \theta} -  2 g^{ii} \Gamma_{i \theta}^{i} \partial_i h_{i \theta} - g^{ii} \Gamma_{ii}^{\theta}
\partial_{\theta}h_{\theta \theta} + \right. \nonumber \\ && \left. g^{ii} \Gamma_{ii}^{\theta} (\frac{A'}{A}) h_{\theta \theta}
-2 g^{ii} \Gamma_{i\theta}^{i}\partial_{i}h_{i \theta} + 2 g^{ii} \Gamma_{i\theta}^{i} \Gamma_{ii}^{\theta}h_{\theta \theta}
+ 2 g^{ii} ( \Gamma_{i\theta}^{i} )^2 h_{ii}  \right ].~~~~~~
\label{a7}
\end{eqnarray}
Suppose we now consider perturbations for which all components are zero except $h_{\theta i }$ ($i$ is either $\chi_1$ or $\chi_2$ ) and $h_{\theta i }$ is a function only of $\theta$ and $\tau$, then substituting in the above formula,
$(\Delta h)_{\theta \theta} = 0$. From (\ref{a3}), we can also conclude that in this case, $(\Delta_L h)_{\theta \theta} = 0$ as well.\\

2. For the formula to follow, $i\neq j \neq \theta$ and there is no summation over repeated indices.
\begin{eqnarray}
&&(\Delta h)_{ii} = g^{\theta \theta} \partial_{\theta}^{2} h_{ii} - 2 g^{\theta \theta} \partial_{\theta} (\Gamma_{\theta i}^{i} h_{ii}) - g^{\theta \theta} \Gamma_{\theta \theta}^{\theta} \left [ \partial_{\theta} h_{ii} - 2 \Gamma_{\theta i}^{i} h_{ii} \right ] -
2 g^{\theta \theta} \Gamma_{\theta i}^{i} \left [ \partial_{\theta} h_{ii} - 2 \Gamma_{\theta i}^{i} h_{ii} \right ]  \nonumber \\&&+ g^{ii} \left [ \partial_{i}^{2} h_{ii} - 2 \partial_{i} (\Gamma_{ii}^{\theta} h_{\theta i}) - \Gamma_{ii}^{\theta} \partial_{\theta} h_{ii} + 2 \Gamma_{ii}^{\theta} \Gamma_{\theta i}^{i} h_{ii} - 2 \Gamma_{ii}^{\theta} \partial_i h_{\theta i} + 2 \left (\Gamma_{ii}^{\theta} \right )^2 h_{\theta \theta} + 2 \Gamma_{ii}^{\theta} \Gamma_{\theta i}^{i} h_{ii} \right ]
 \nonumber \\ &&+ g^{jj} \left [ \partial_{j}^{2} h_{ii} - \Gamma_{jj}^{\theta} (\partial_{\theta} h_{ii} - 2 \Gamma_{\theta i}^{i} h_{ii}) \right ].
\label{a8}
\end{eqnarray}
We again observe that for the class of perturbations of the sausage where the only nonzero components are $h_{\theta i}$ which are functions only of $\theta$ and $\tau$, $(\Delta_L h)_{ii} = 0$.\\

3. Let $i\neq j \neq \theta$ and there is no summation over repeated indices. Then
\begin{eqnarray}
&&(\Delta h)_{ij} =  g^{\theta \theta} \left [ \partial_{\theta}^{2} h_{ij } - \partial_{\theta}(\Gamma_{\theta i}^{i} h_{ij}) - \partial_{\theta}(\Gamma_{\theta j}^{j} h_{ij}) - (\Gamma_{\theta \theta}^{\theta} + \Gamma_{\theta i}^{i} + \Gamma_{\theta j}^{j}) \left ( \partial_{\theta}h_{ij} - \Gamma_{\theta i}^{i} h_{ij} - \Gamma_{\theta j}^{j} h_{ij} \right ) \right ] \nonumber \\
&&+ g^{ii} \left[ \partial_{i}^{2} h_{ij } - \partial_i (\Gamma_{ii}^{\theta} h_{\theta j}) - \Gamma_{ii}^{\theta} \left (\partial_{\theta} h_{ij} - \Gamma_{\theta i}^{i} h_{ij} - \Gamma_{\theta j}^{j} h_{ij} \right) - \Gamma_{ii}^{\theta} \left (\partial_{i} h_{\theta j} - \Gamma_{\theta i}^{i} h_{ij} \right ) \right ] \nonumber \\ &&+ g^{jj} \left[ \partial_{j}^{2} h_{ij } - \partial_j (\Gamma_{jj}^{\theta} h_{\theta i}) - \Gamma_{jj}^{\theta} \left (\partial_{\theta} h_{ij} - \Gamma_{\theta j}^{j} h_{ij} - \Gamma_{\theta i}^{i} h_{ij} \right) - \Gamma_{jj}^{\theta} \left (\partial_{j} h_{\theta i} - \Gamma_{\theta j}^{j} h_{ij} \right ) \right ].~~~~~~~~~~~~~~~
\label{a9}
\end{eqnarray}
For the special class of perturbations discussed below (\ref{a8}), $(\Delta_L h)_{ij} = 0$.\\

4. Let $i\neq j \neq \theta$ and there is no summation over repeated indices.
\begin{eqnarray}
&&(\Delta h)_{\theta i} = g^{\theta \theta} \left [ \partial_{\theta}^{2} h_{\theta i } - \partial_{\theta} (\Gamma_{\theta \theta}^{\theta} h_{\theta i }) - \partial_{\theta}(\Gamma_{\theta i}^{i} h_{\theta i}) \right ] - 2 g^{\theta \theta} \Gamma_{\theta \theta}^{\theta} \left [ \partial_{\theta}h_{\theta i } - \Gamma_{\theta \theta}^{\theta}h_{\theta i } - \Gamma_{\theta i}^{i}h_{\theta i } \right ] \nonumber \\ &&- g^{\theta \theta} \Gamma_{\theta i}^{i} \left [ \partial_{\theta}h_{\theta i } - \Gamma_{\theta \theta}^{\theta}h_{\theta i } - \Gamma_{\theta i}^{i}h_{\theta i } \right ] + g^{ii} \left [ \partial_{i}^{2} h_{\theta i } - \partial_{i}(\Gamma_{ii}^{\theta} h_{\theta \theta}) - \partial_{i}(\Gamma_{\theta i}^{i} h_{ii} ) \right ] \nonumber \\ && - g^{ii} \Gamma_{ii}^{\theta} \left [ \partial_{\theta} h_{\theta i } - \Gamma_{\theta \theta}^{\theta} h_{\theta i } - \Gamma_{\theta i}^{i} h_{\theta i} \right ] - g^{ii} \Gamma_{\theta i}^{i} \left [\partial_i h_{ii} - 2 \Gamma_{ii}^{\theta} h_{\theta i } \right ] - g^{ii} \Gamma_{ii}^{\theta} \left [ \partial_i h_{\theta \theta} - 2 \Gamma_{\theta i}^{i}h_{\theta i} \right ] \nonumber \\ &&+ g^{jj} \left [ \partial_{j}^{2} h_{\theta i } - \partial_{j}(\Gamma_{\theta j}^{j} h_{ij} ) \right ] - g^{jj} \Gamma_{jj}^{\theta} \left [ \partial_{\theta} h_{\theta i } - \Gamma_{\theta \theta}^{\theta} h_{\theta i } - \Gamma_{\theta i}^{i} h_{\theta i} \right ] \nonumber \\ && - g^{jj} \Gamma_{\theta j}^{j} \left [ \partial_j h_{ij} - \Gamma_{jj}^{\theta}h_{\theta i } \right ].
\label{a10}
\end{eqnarray}
For the special class of perturbations discussed below (\ref{a8}), $(\Delta_L h)_{\theta i} \neq 0$. Therefore, a restriction to this special class of perturbations is possible under the flow (\ref{4.5}) --- in the sense that there are nontrivial perturbations in this special class which solve (\ref{4.5}).

\end{document}